\documentclass[iop]{emulateapj}
\usepackage{epsfig}

\journalinfo{ApJ Letters}
\slugcomment{ApJL accepted}

\shortauthors{KERR ET AL.}
\shorttitle{5 Millisecond Pulsars in Fermi Sources}

\begin{document}
%\linenumbers

%  Definitions
\def\fermi{{\em Fermi}}
\def\ergs{\,erg\,s$^{-1}$}
\def\ergcmsqs{\,erg\,\,cm$^{-2}$\,s$^{-1}$}
\def\pccc{pc\,cm$^{-3}$}

% Objects
%\def\fgl{1FGL~J2030.0+3641}
\def\psr{PSR~J0101--6422}

% Shortcuts
\def\hr{^{\rm h}}
\def\mi{^{\rm m}}
\def\msun{M_{\sun}}

\title{Five New Millisecond Pulsars From a Radio Survey of 14 
Unidentified Fermi-LAT Gamma-ray Sources}
\author{
M.~Kerr\altaffilmark{1,2,3}, 
F.~Camilo\altaffilmark{4,5}, 
T.~J.~Johnson\altaffilmark{6,7}, 
E.~C.~Ferrara\altaffilmark{8}, 
L.~Guillemot\altaffilmark{9}, 
A.~K.~Harding\altaffilmark{8}, 
J.~Hessels\altaffilmark{10,11}, 
S.~Johnston\altaffilmark{12}, 
M.~Keith\altaffilmark{12}, 
M.~Kramer\altaffilmark{13,9}
S.~M.~Ransom\altaffilmark{14}, 
P.~S.~Ray\altaffilmark{15}
J.~E.~Reynolds\altaffilmark{12},
J.~Sarkissian\altaffilmark{16},
and K.~S.~Wood\altaffilmark{15}
}
\altaffiltext{1}{W. W. Hansen Experimental Physics Laboratory, Kavli Institute for Particle Astrophysics and Cosmology, Department of Physics and SLAC National Accelerator Laboratory, Stanford University, Stanford, CA 94305, USA}
\altaffiltext{2}{email: kerrm@stanford.edu}
\altaffiltext{3}{Einstein Fellow}
\altaffiltext{4}{Columbia Astrophysics Laboratory, Columbia University, New York, NY 10027, USA}
\altaffiltext{5}{email: fernando@astro.columbia.edu}
\altaffiltext{6}{National Research Council Research Associate, National Academy of Sciences, Washington, DC 20001, resident at Naval Research Laboratory, Washington, DC 20375, USA}
\altaffiltext{7}{email: tyrel.j.johnson@gmail.com}
\altaffiltext{8}{NASA Goddard Space Flight Center, Greenbelt, MD 20771, USA}
\altaffiltext{9}{Max-Planck-Institut f\"ur Radioastronomie, Auf dem H\"ugel 69, 53121 Bonn, Germany}
\altaffiltext{10}{Netherlands Institute for Radio Astronomy (ASTRON),Postbus 2, 7990 AA Dwingeloo, Netherlands}
\altaffiltext{11}{Astronomical Institute ``Anton Pannekoek'' University of Amsterdam, Postbus 94249 1090 GE Amsterdam, Netherlands}
\altaffiltext{12}{CSIRO Astronomy and Space Science, Australia Telescope National Facility, Epping NSW 1710, Australia}
\altaffiltext{14}{National Radio Astronomy Observatory (NRAO), Charlottesville, VA 22903, USA}
\altaffiltext{13}{Jodrell Bank Centre for Astrophysics, School of Physics and Astronomy, The University of Manchester, M13 9PL, UK}
\altaffiltext{15}{Space Science Division, Naval Research Laboratory, Washington, DC 20375-5352, USA}
\altaffiltext{16}{CSIRO Parkes Observatory, Parkes NSW 2870, Australia}
%\linenumbers
\begin{abstract}
We have discovered five millisecond pulsars (MSPs) in a survey of 14
unidentified \fermi-LAT sources in the southern sky using the Parkes
radio telescope.  PSRs J0101--6422, J1514--4946, and J1902--5105
reside in binaries, while PSRs J1658--5324 and J1747--4036 are
isolated.  Using an ephemeris derived from timing
observations of \psr{} ($P$=2.57\,ms, DM=12\,\pccc), we have detected
$\gamma$-ray pulsations and measured its proper motion.  Its
$\gamma$-ray spectrum (a power law of $\Gamma=0.9$ with a cutoff at
1.6\,GeV) and efficiency are typical of other MSPs, but its radio and
$\gamma$-ray light curves challenge simple geometric models of
emission.  The high success rate of this survey---enabled by selecting
$\gamma$-ray sources based on their detailed spectral
characteristics---and other similarly successful searches indicate
that a substantial fraction of the local population of MSPs may soon
be known.
\end{abstract}

\section{Introduction} \label{sec:intro} 

The Large Area Telescope \citep[LAT,][]{lat_instrument} of the
\textit{Fermi Gamma-ray Space Telescope} is a pulsar detector {\it par
excellence}.  Supported by a radio campaign
\citep{smith_timing} providing ephemerides for coherent folding of LAT
photons, the LAT has identified
$\gamma$-ray pulsations from many normal (unrecycled) pulsars
\citep[e.g.][]{psr_cat} and millisecond pulsars
\citep[MSPs,][]{msp_pop}.  Additionally, tens of new pulsars have been
discovered in ``blind'' searches for periodicity in the LAT data
\citep[e.g.][]{blind_search_16,pletsch+_2012}.

A third, indirect method of detection has been extremely successful in
discovering new pulsars, especially MSPs\footnote{See
https://confluence.slac.stanford.edu/display/GLAMCOG/Public+List+of+LAT-Detected+Gamma-Ray+Pulsars
for an up-to-date list of LAT-detected pulsars}.  Gamma-ray
pulsars---both young and recycled---are stable emitters and nearly all
have a characteristic spectrum $dN/dE\propto E^{-\Gamma}\exp(-E/E_c)$
with $\Gamma < 2$ and $1 < E_c/\mathrm{GeV} < 10$ \citep{psr_cat},
making unidentified, nonvariable LAT sources with such spectra good
pulsar candidates.  The fine angular resolution and large effective
area of the LAT allow a typical source localization of
$\sim$10\arcmin, about the $1\,{\rm GHz}$ beam size of $100\,{\rm m}$
class radio telescopes.  This happy coincidence enables deep pulsation
searches with a single pointing.  Discovery of a pulsar and a
subsequent timing campaign can ``close the loop'' by providing an
ephemeris with which to fold LAT photons and resolve $\gamma$-ray
pulsations.  Indeed, since most MSPs reside in binaries, initial
detection and characterization of the orbit at longer wavelengths is
the \emph{only} way $\gamma$-ray pulsations can be detected from these
pulsars.
% \citep[c.f.][where coverage of an
%EGRET error box required 52 beams!]{crawford+_2006}

Due to sensitivity limitations, LAT-guided surveys naturally target
the nearly-isotropic population of nearby MSPs.  The deep radio
exposures afforded by the efficient target selection offer the
tantalizing possibility of completely cataloguing the local field MSPs
whose radio and $\gamma$-ray beams cross the Earth.  This relatively
complete sample will be valuable in constraining the emission
mechanisms of MSPs and in characterizing the evolution of their binary
progenitors.

Towards this end, we performed a small, targeted survey at the CSIRO
Parkes telescope, complementary to a Parkes search described in
\citet{keith+_2011} that had different source selection criteria.

\section{Target Selection} \label{sec:targets}

We began with the set of southern (Decl. $<-40\arcdeg$; northern
sources are visible to the more sensitive Green Bank Telescope),
nonvariable sources in a preliminary version of the 1FGL catalog of
$\gamma$-ray sources \citep{1fgl}.  We further restricted
consideration to sources whose LAT position estimates had a 95\%
confidence error radius $\leq$7\arcmin, the HWHM of the 1.4\,GHz beam
of the Parkes telescope, and which had no plausible blazar
counterpart.  We classified the remaining sources by visual inspection
of the $\gamma$-ray spectrum.  Sources with spectral shapes resembling
known pulsars (see above) were selected for observation, while those
sources best described by power laws or with significant spectral
breaks but $\Gamma>2$ were deemed likely to be blazars and were not
selected.  Due to the difficulty of spectral modeling and localization
of LAT sources in the Galactic plane, we discarded sources with
$|b|<5\arcdeg$.  Our final list comprised 14 good candidates.  The
source positions (see Table~1) were computed from the same analysis
used to estimate the spectral shape, but in all cases the difference
in position with 1FGL is well within both the 1FGL 95\% error contour
and the Parkes beam.

\section{Radio Searches}\label{sec:survey} 

We observed the 14 selected sources (\S\ref{sec:targets}) at the
Parkes 64-m radio telescope between 2009 November 25 and December 8.
Each source was observed for between 1 and 2 hours (see Table~1) using
the center beam of the multibeam receiver.  We recorded
1-bit-digitized total-power samples every 125\,$\mu$s from each of 512
frequency channels spanning 256\,MHz of band centered on 1390\,MHz,
writing the data to disk for off-line analysis (see
\citet{manchester+_2001_parkes_multibeam} for
more details on the receiver and data-acquisition system.)

The data were analyzed on the {\tt koala} computer cluster at Columbia
University using standard pulsar search techniques implemented in the
PRESTO package \citep{ransom_thesis}.  We dedispersed each data set
ideally up to a dispersion measure of $\mbox{DM}=270$\pccc,
and when searching for periodic signals we allowed for accelerated
signals caused by pulsar motion in a binary system.  This was
parameterized within PRESTO by the parameter ${\tt zmax}=50$, which is
related to the maximum number of bins a signal can drift within the
Fourier spectrum and still be properly detected by the search method.
All of the data were analyzed in this manner within about a week of
collection, and on 2009 December 4 we found our first MSP.  Within a
few more days we confirmed it and four other new MSPs.  A 6th MSP was
detected unbiasedly in our sample, but it had been previously
discovered by \citet[][see Table~1]{keith+_2011}.

In a few cases, the pulsars were not confirmed on the first attempt
and required subsequent observations, owing to their
faintness and scintillation in the interstellar medium.  We reanalyzed
the data by increasing the acceleration search space to ${\tt
zmax}=200$ and by analyzing only the first 30 minutes of each data
file.  The latter approach improves our sensitivity to highly
accelerated pulsars (e.g., for an MSP in a 1-day binary, a 2-hour
observation cannot be corrected ideally under the assumption of a
constant acceleration, no matter how large the {\tt zmax} used), but
no new pulsars were detected.

The nominal sensitivity of our searches, for a 2\,hr observation and
assuming a pulsar duty cycle of 20\%, is 0.11\,mJy for $P\ga$2\,ms and
$\mbox{DM}\la 40$\,pc\,cm$^{-3}$, degrading gradually for shorter
periods and larger DMs.  While in detail this depends on the sky
background temperature, the contribution from the Galaxy at our
observing frequency only differs by a maximum of 3\,K for different
locations, out of a total system equivalent temperature of nearly
30\,K, so that we provide an average flux density limit.  Two of
the non-detections have limits of 0.12\,mJy and 0.14\,mJy owing to
shorter integration times (see Table~\ref{tab:survey}).  Because of
scintillation, a pulsar with a larger average flux than our nominal
limits might not be detectable in a particular observation, and vice
versa.  In any case, the nominal flux density represents the most
sensitive searches for MSPs done at Parkes, especially considering
that these LAT-selected targets are largely expected to be relatively
nearby.

%We began timing observations of the new pulsars in 2009 December.
Characterization of four of the MSPs requires a longer radio timing
campaign and/or LAT dataset than available for this work.  We defer
discussion of these MSPs (and of scintillation effects on sensitivity,
which likewise requires repeated observations) to Camilo et al. (in
preparation).  The remaining pulsar presents a peculiar light
curve and we describe it further below.

\begin{deluxetable*}{lcccrccccccl}
%\centering
\tablewidth{0.99\linewidth}
\tablecaption{\label{tab:survey} Results of Radio Searches of 14 1FGL Sources at Parkes }
\tablecolumns{12}
\tablehead{
\colhead{1FGL Name} &
\colhead{R.A.\tablenotemark{a}} &
\colhead{Decl.\tablenotemark{a}} &
\colhead{l} &
\colhead{b} &
\colhead{$\Gamma$} &
\colhead{$\mathrm{E_c}$} &
\colhead{Obs. Time} &
\colhead{Period} &
\colhead{DM} &
\colhead{Distance\tablenotemark{b}} &
\colhead{Binary} \\
\colhead{} &
\colhead{(J2000.0)} &
\colhead{(J2000.0)} &
\colhead{(deg)} &
\colhead{(deg)} &
\colhead{} &
\colhead{(GeV)} &
\colhead{(hr)} &
\colhead{(ms)} &
\colhead{(\pccc)} &
\colhead{(kpc)} &
\colhead{}
}
\startdata
J0101.0--6423 & $01\hr00\mi58\fs1$ & $-64\arcdeg24'03\arcsec$ & 301.2 & --52.7 & 1.3 & 2.3 & 1.0  & 2.57& 11.93 & 0.55 & yes \\
J0603.0--4012 & $06\hr03\mi04\fs9$ & $-40\arcdeg11'02\arcsec$ & 246.8 & --25.9 & 1.9 & 6.5 & 2.0  & --- & ---   & ---  & --- \\
J0933.9--6228 & $09\hr33\mi58\fs5$ & $-62\arcdeg27'54\arcsec$ & 282.2 & --7.8  & 0.1 & 1.2 & 2.0  & --- & ---   & ---  & --- \\
J1036.2--6719 & $10\hr36\mi16\fs9$ & $-67\arcdeg20'34\arcsec$ & 290.4 & --7.8  & 1.5 & 2.2 & 2.0  & --- & ---   & ---  & --- \\
J1227.9--4852 & $12\hr27\mi50\fs5$ & $-48\arcdeg51'54\arcsec$ & 298.9 & 13.8  & 1.8 & 1.6 & 2.0  & --- & ---   & ---  & --- \\
J1232.2--5118 & $12\hr31\mi49\fs1$ & $-51\arcdeg18'50\arcsec$ & 299.8 & 11.4  & 1.8 & 2.3 & 2.0  & --- & ---   & ---  & --- \\
J1514.1--4945 & $15\hr14\mi05\fs7$ & $-49\arcdeg45'32\arcsec$ & 325.2 & 6.8   & 1.7 & 5.9 & 2.0  & 3.59& 31.5  & 0.9 & yes \\
J1624.0--4041 & $16\hr24\mi06\fs2$ & $-40\arcdeg40'48\arcsec$ & 340.6 & 6.2   & 2.1 & 3.3 & 2.0  & --- & ---   & ---  & --- \\
J1658.8--5317 & $16\hr58\mi43\fs2$ & $-53\arcdeg17'45\arcsec$ & 335.0 & --6.6  & 2.1 & 1.8 & 1.3  & 2.44& 30.9  & 0.9 & no  \\
J1743.8--7620 & $17\hr43\mi44\fs6$ & $-76\arcdeg20'42\arcsec$ & 317.1 & --22.5 & 1.2 & 2.0 & 1.5  & --- & ---   & ---  & --- \\
J1747.4--4035 & $17\hr47\mi29\fs1$ & $-40\arcdeg36'07\arcsec$ & 350.2 & -6.4  & 1.5 & 3.4 & 1.4  & 1.65& 152.9 & 3.3  & no  \\
J1902.0--5110 & $19\hr02\mi05\fs5$ & $-51\arcdeg09'43\arcsec$ & 345.6 & --22.4 & 1.7 & 4.4 & 1.2  & 1.74& 36.3  & 1.2  & yes \\
J2039.4--5621 & $20\hr39\mi30\fs5$ & $-56\arcdeg20'42\arcsec$ & 341.2 & -37.1 & 1.6 & 2.7 & 1.2  & --- & ---   & ---  & --- \\
J2241.9--5236\tablenotemark{c} & $22\hr41\mi52\fs4$ &$-52\arcdeg37'37\arcsec$ &337.4 & -54.9 & 1.6 & 3.6 & 2.0  & 2.19& 11 & 0.5 & yes \\[-5pt]
\enddata
%\tablecomments{Stub for any comments.}
\tablenotetext{a} {Parkes telescope pointing position.}
\tablenotetext{b} {Distances computed from measured DM using the NE2001 model \citep{cl02}.}
\tablenotetext{c} {See \citet{keith+_2011}.}
\end{deluxetable*}

\section{\psr{}}
\subsection{Timing Solution} \label{sec:timing}

After the discovery observations on 2009 November 25 and confirmation
on December 8, we began regular timing observations of \psr\ at
Parkes.  We observed the pulsar on 35 days through 2011 November 9,
detecting it on 28 occasions; on the remaining 7 days it was too faint
to detect, due to interstellar scintillation (\mbox{DM}=12\,\pccc.)
Each observation was typically one hour, with the same receiver
and data acquisition system used in the search observations.  Using
TEMPO2\footnote{http://www.atnf.csiro.au/research/pulsar/tempo2} with
the 28 times-of-arrival, we obtained a phase-connected timing
solution whose parameters are given in Table~\ref{tab:parms}.

% Additionally, using the aperture described in \S\ref{sec:gamprof}, we
% extracted 9 TOAs from the 36 months of LAT data
% \citep{ray+_pulsar_timing}.  Despite the lower precision
% ($\sim$60\,$\mu$s vs. 2--20\,$\mu$s), the extended timing baseline
% allows an improved detection of the pulsar's proper motion,
% $21\pm3$\,mas\,yr$^{-1}$.  The parameters of this solution, derived
% from both radio and $\gamma$ TOAS, are listed in
% Table~\ref{tab:parms}.

The Shklovskii effect \citep{shklovskii} increases the measured value
of $\dot{P}=5.2\times10^{-21}$ beyond that intrinsic to the pulsar by
$\Delta \dot{P}=P\,v_{\perp}^2/d\,c$, with $v_{\perp}$ the source
velocity transverse to the line-of-sight.  For nearby objects, it may
dominate the true spindown rate and lead to overestimates of
the spindown luminosity \citep{camilo+_1994}.  The observed proper
motion, at the 0.55\,kpc DM distance \citep{cl02}, implies a true
$\dot{P}_i=(4.3\pm0.1)\times10^{-21}$.  At this distance, the
indicated $v_{\perp}=41$ km\,s$^{-1}$ is typical for an MSP
\citep[see][]{nice&taylor_1995}.%  Continued timing will be useful
%either to limit or (optimistically) to measure the pulsar's parallax,
%for if the pulsar is much closer than its DM distance, the
%contribution of the Shklovskii effect is substantially larger than
%given here.

%For an accretion rate of
%$\sim$3$\times10^{-10}$ $\msun$ yr (e.g., 0.8\,$\msun$ over 3 Gyr), a
%magnetic field of $10^8$G, and a NS radius of 10$^6$\,cm, the
%equilibrium MSP spin period \citep[e.g. Eq. 2.10
%of][]{bhattcharya&van_den_heuvel_1991} is about 1.8\,ms.  Even with
%the relatively low $\dot{P}_i$ of this pulsar, the current 2.6\,ms
%period can be reached in a pulsar lifetime of 10 Gyr.  On the other
%hand, if the mass of the donor star was close to 1\,$\msun$ and the
%pulsar is only a few Gyr old, the accretion rate (initial spin period)
%may have been appreciably lower (higher).

\subsection{Radio Profile and Polarimetry} \label{sec:pol}
\begin{figure}[t]
\centerline{
\hfill
\epsfig{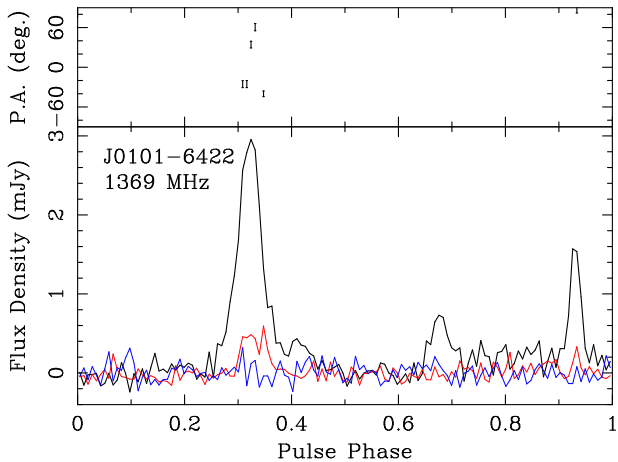}
\hfill
}
\caption{\label{fig:pol}
Polarimetric pulse profile of \psr\ at 1.4\,GHz, displayed with 128
phase bins, based on 6\,hr of Parkes PDFB3 data using 256\,MHz of
bandwidth.  The black trace corresponds to total intensity, red to linear
polarization, blue to circular.  In the upper window, the position angle
of linear polarization is plotted for bins with linear signal-to-noise
ratio $>3$.  The pulse peak is displayed with arbitrary phase, and the
mean flux density is $\sim0.2$\,mJy.
}
\end{figure}

We made polarimetric and flux-calibrated observations of \psr,
using the PDFB3 digital filterbank.  The data were analyzed with
PSRCHIVE \citep{hotan&manchester_2004}.  The best full-Stokes profile
that we obtained is shown in Figure~\ref{fig:pol} (the flux density of
this profile is not necessarily representative of the average pulsar
intensity, which varies greatly owing to interstellar scintillation).
The main pulse is linearly polarized at the $\sim$15\% level, but the
low signal-to-noise ratio prevents a useful determination of rotation
measure (the best-fit value is $\sim +10$\,rad\,m$^{-2}$, but within
the uncertainties is consistent with zero).  These data show that the
subsidiary pulse component is composed of two outer peaks joined by a low bridge of radio emission.

\subsection{Gamma-ray Profile} \label{sec:gamprof}

\begin{figure}[t]
\centerline{
\hfill
\epsfig{file=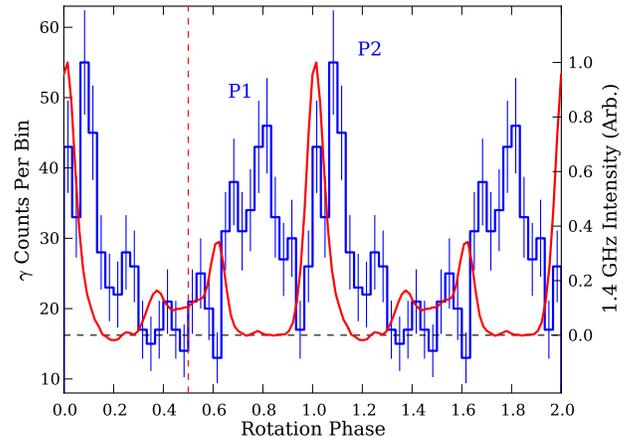,width=0.99\linewidth,clip=}
\hfill
}
\caption{\label{fig:prof}
Phase-aligned $\gamma$-ray (blue histogram) and radio (red trace)
pulse profiles of \psr{}.  The radio profile is summed from 28 timing
detections, while the $\gamma$-ray profile corresponds to the optimal
aperture described in the main text.  The background estimation
(dashed horizontal line) is derived by counting photons in the
off-pulse phase window (\S\ref{sec:spect}) and is consistent with the
off-pulse background level obtained via spectral analysis.  The
vertical dashed line indicates the phase ($\phi\equiv0.5$) from which
the offset of ``P1'' is measured.
}
\end{figure}

\begin{figure}[t]
\centerline{
\hfill
\epsfig{file=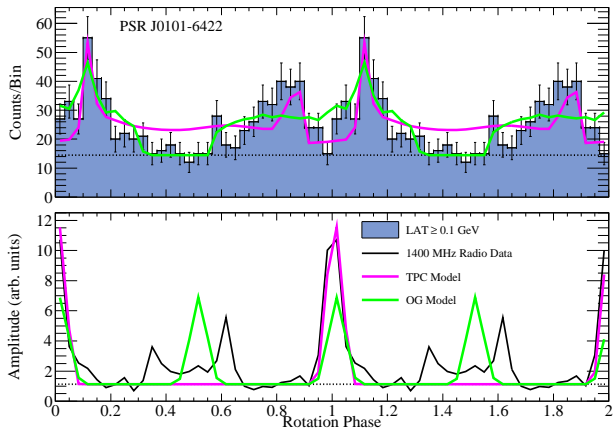,width=0.99\linewidth,clip=}
\hfill
}
\caption{\label{fig:lcmodel}
The best-fit model light curves for the two-pole caustic (TPC) and
outer gap (OG) models (see \S\ref{subsec:disc_psr}.)  The best fit
geometries ($\alpha$,$\,\zeta$) for TPC and OG are
(26$\arcdeg$,$\,79\arcdeg$) and (90$\arcdeg$,$\,36\arcdeg$).
}
\end{figure}

To characterize the $\gamma$-ray profile, we selected photons
collected between 2008 Aug 4 and 2011 Aug 1 with reconstructed
energies 0.3--3\,GeV lying within 1.1\arcdeg\  of the timing position.
This selection balances completeness (including as
many pulsed photons as possible) and pulsed signal-to-noise ratio as measured by the
H-test \citep{dejager_1}.  %We note that while pulsed emission
%continues up to 10\,GeV, the 3--10\,GeV band contributes only a few
%photons and we exclude it from the profile analysis.
We apply the
same data processing as described in \S\ref{sec:spect}, omitting the
horizon cut to increase the livetime by about 20\%, at the expense
of a slightly increased background.

The light curve corresponding to this extraction appears in Figure
\ref{fig:prof}.  If $\gamma$ rays are produced above the null charge
surface (the locus of points where the magnetic field is orthogonal to
the pulsar spin axis), they appear to an observer in the {\em
opposite} hemisphere, whereas radio emission from low altitudes is
beamed into the same hemisphere
\citep[see][]{romani&yadigaroglu_1995}.  In this scenario, the
line-of-sight to \psr{} would intercept a cone of radio emission in
one hemisphere to produce the weaker radio peak at $\phi\sim0.5$ and a
cone of radio and $\gamma$ emission in the other to produce the two
$\gamma$ peaks and bright radio peak at $\phi\sim1$.  In keeping with
this picture and the convention established in the literature, we
identify the $\gamma$ peak at $\phi\sim0.8$ as ``P1'' and that at
$\phi\sim0.1$ as ``P2''.%  The
%small feature at $\phi\sim0.6$ is not statistically significant but
%may become interesting as additional data are collected.

The relative phasing of the peaks is of interest
\citep[e.g.][]{watters+_2009}. The radio-to-$\gamma$ offset, $\delta$,
is measured from the center of the leading radio peak (taken as
$\phi\equiv0.5$, see Figure \ref{fig:prof}).  We estimated the phase
of the $\gamma$-ray peaks using unbinned maximum likelihood with a
two-sided Lorentzian model for the peaks.  These estimates yield
$\delta=0.33\pm0.01_{\rm stat}\pm0.03_{\rm syst}$ and a $\gamma$-ray peak-to-peak separation
of $\Delta=0.26\pm0.03_{\rm stat}\pm0.02_{\rm syst}$.  The DM-induced uncertainty in
$\delta$ is $<0.001P$.  The systematic uncertainty for both
$\delta$ and $\Delta$ is estimated from the scatter obtained by
fitting alternative functional forms for the peaks.
%and.%, primarily for $\delta$, from the breadth of the radio peaks.

% uncertainty in DM of 0.001 @ 1.4 GHz --> 2.1mus of uncertainty
% this is only 0.0008P

%Note there is no evidence for emission below 300 MeV.
%Actually, there is now very modest (2-3 sigma) evidence for emission
%below 300 MeV and above 3 GeV

\subsection{Gamma-ray Spectrum} \label{sec:spect}

To measure the spectrum of psr{}, we analyzed low background {\tt
DIFFUSE} class ``Pass 6'' \citep{lat_instrument} LAT data collected
between 2008 Aug 4 and 2011 Apr 17.  We filtered periods where the
observatory's rocking angle exceeded $52\arcdeg$ or the Earth's
limb impinged upon the field of view (horizon cut, requires zenith
angle $<100$\arcdeg).  We modelled the sensitivity to these events
with the flight-corrected {\tt P6\_V11\_DIFFUSE} instrument response
function.  To compute background contributions, we used a preliminary version of the 2FGL catalog and its
accompanying diffuse emission
models\footnote{http://fermi.gsfc.nasa.gov/ssc/data/access/lat/BackgroundModels.html}.  We selected events within
$10\arcdeg$ and re-fit the spectra of point sources within $8\arcdeg$
and the normalizations of the diffuse models.  Fits employed {\it
pointlike} \citep{kerr_thesis}, a binned maximum likelihood algorithm.

The phase-averaged spectrum of \psr{} is well described by an
exponentially cutoff power law, $dN/dE =
N_0\,(E/E_0)^{-\Gamma}\,\exp(-E/E_c)$, whose parameters appear in
Table \ref{tab:parms}.  We also considered separately the spectra of
the three major components, viz.  P2 ($0 < \phi < 0.33$), the
off-pulse ($0.33<\phi<0.63$), and P1 ($0.63 < \phi < 1$).  The
parameter uncertainties are relatively large, and accordingly we found
no significant difference in the spectral shapes of the
two peaks.  Further, we detected no significant emission in the
off-pulse phase window and derived a $95\%$ confidence upper limit on
the energy flux of a point source with a power law spectrum with
$\Gamma=2.2$, obtaining $6\times10^{-12}$ erg cm$^{-2}$ s$^{-1}$
($>100$ MeV; scaled up by 3.3 to the full phase window.)

The observed flux is related to the $\gamma$-ray luminosity by
$F_{\gamma} = L_{\gamma}/4\pi\,f_{\Omega}\,d^2$, where $f_{\Omega}$ is
an unknown beaming factor \citep[typically of order unity,
e.g.][]{watters+_2009} implying a phase-averaged, isotropic
($f_{\Omega}=1$) $\gamma$-ray luminosity $L_{\gamma}=
4.0\times10^{32}$\ergs$ = 0.04\,\dot{E}$ at $d=0.55$ kpc.
Best-fit values of $f_{\Omega}$ from the TPC and OG models
(\S\ref{subsec:disc_psr}) are 0.7 and 0.9. This $\gamma$-ray
efficiency is in accord with other LAT-detected MSPs \citep{msp_pop}.

\subsection{X-ray and Optical Observations}

%WebPimms gives
%PIMMS predicts 2.300E-03 cps with SWIFT XRT PC ( 0.500- 8.000keV)
%(Grade 0-12 on-axis count rate for an infinite extraction region)
%or 1.781E-03 cps in Grade 0 
% for \Gamma=1.5, N_H = 3.7e20, Flux (0.5-8 kev) = 1e-13 erg/cm2/s.

% Implies 1.12e-13 erg/cm2/s upper limit
% Luminosity at 0.55kpc = 4.08e30 erg/s

A 3.3\,ks \textit{Swift} observation with the X-ray Telescope in PC
mode was obtained on 2009 Nov 21 to search for an X-ray counterpart to
the then-unidentified 1FGL source.  No significant counterpart is
detected, with a 3$\sigma$ upper limit at the pulsar position of
$2.0\times10^{-3}$\,cts\,s$^{-1}$ (0.5--8 keV).  Assuming a power-law
spectrum with photon index $\Gamma=1.5$ for a column with
$N_{\mathrm{H}} = 3.7\times10^{20}$\,cm$^{-2}$ (10 H{\sc I} atoms per
free electron along the line-of-sight) yields an unabsorbed flux limit
of $1.1\times10^{-13}$\ergcmsqs, or efficiency
$L_X/\dot{E}<5.0\times10^{-4}$ at 0.55\,kpc.  This limit is comparable
to the detected X-ray flux from other MSPs
\citep[e.g.][]{marelli+_2011}.

We searched four DSS plates and found no optical counterpart
consistent with the pulsar's position, suggesting $m>21$.

%At 0.55kpc, this
%corresponds to $L_X/\dot E < X\times10^{-3}$, which is a reasonably
%stringent limit, although it would not be a surprise if the actual
%X-ray flux of this pulsar were about an order of magnitude below the
%current limit \citep[see, e.g.,][]{mdc11}.

\begin{deluxetable*}{ll}
\tablewidth{0.61\linewidth}
\tablecaption{\label{tab:parms} Measured and Derived Parameters
for \psr{}}
\tablecolumns{2}
\tablehead{
\colhead{Parameter} &
\colhead{Value}
}
\startdata
Right ascension, R.A. (J2000.0)\dotfill & $01^{\rm h}01^{\rm m}11\fs1163(2)$  \\
Declination, decl. (J2000.0)\dotfill    & $-64\arcdeg22'30\farcs171(2)$       \\
% NB dRA/dt = PMRA/cos(decl) = 10.226 / 0.432
Proper motion in R.A. $\times\cos$ decl. (mas yr$^{-1}$)\dotfill  & 10(1)  \\
Proper motion in decl. (mas yr$^{-1}$)\dotfill & --12(2)                        \\
Galactic longitude, $l$ (deg.)\dotfill  & 301.19                              \\
Galactic latitude, $b$ (deg.)\dotfill   & --52.72                             \\
Spin period, $P$ (ms)\dotfill           & 2.5731519721683(2)                  \\
Period derivative (apparent), $\dot P$\dotfill     & $5.16(3)\times10^{-21}$            \\
%Period derivative (intrinsic), $\dot{P}_i$\dotfill     & $4.1(1)\times10^{-21}$            \\
Epoch of period (MJD)\dotfill           & 55520                               \\
Orbital period, $P_b$ (days)\dotfill     & 1.787596706(2)                      \\
Epoch of periastron (MJD)\dotfill       & 55162.4011764(3)                    \\
Semi-major axis (lt s)\dotfill           & 1.701046(2)                         \\
Eccentricity\dotfill                    & $<1.5\times10^{-5}$                 \\
Dispersion measure, DM (\pccc)\dotfill  & 11.926(1)                           \\
Spin-down luminosity\tablenotemark{a}, $\dot E$ (${\rm
erg\,s^{-1}}$)\dotfill & $1.0\times10^{34}$ --16\%  \\
Characteristic age\tablenotemark{a}, $\tau_c$ (yr)\dotfill &
$9\times10^9$ +20\%      \\
Surface dipole magnetic field strength
(Gauss)\tablenotemark{a}\dotfill & $1.1\times10^{7}$ --9\%   \\
Mean flux density\tablenotemark{b} at 1.4\,GHz, $S_{1.4}$ (mJy)\dotfill & $0.28 \pm 0.06$ \\
Radio--$\gamma$-ray profile offset, $\delta$ ($P$)\dotfill       &$0.35\pm0.01\pm0.05$ \\
$\gamma$-ray profile peak-to-peak separation, $\Delta$ ($P$)\dotfill & $0.26\pm0.03\pm0.02$ \\
$\gamma$-ray ($>0.2$\,GeV) photon index, $\Gamma$\dotfill  & $0.9\pm0.3\pm0.2$         \\
$\gamma$-ray cut-off energy, $E_c$ (GeV)\dotfill           &$1.6\pm0.4\pm0.3$         \\
Photon flux ($>0.1$\,GeV) ($10^{-9}\,{\rm cm^{-2}\,s^{-1}}$)\dotfill &$9.5\pm1.8^{+1.5}_{-2.0}$ \\
Energy flux ($>0.1$\,GeV), $F_{\gamma}$ ($10^{-11}$\ergcmsqs)\dotfill                & $1.1\pm0.1\pm0.1$ \\[-5pt]
\enddata
\tablecomments{Timing solution parameters given in TDB 
relative to the DE405 planetary ephemeris.  Numbers in parentheses
represent the measured $1\,\sigma$ TEMPO2 timing uncertainties on the
last digits quoted.  For $\gamma$-ray parameters, the first uncertainty
is statistical and the second systematic. }
\tablenotetext{a} {Corrected for Shklovskii effect at 0.55\,kpc; the percentage change from
the nominal value is given.}
\tablenotetext{b} {Average flux for 28 timing detections.}
%\tablecomments{Stub for table comments.}
\end{deluxetable*}

\section{Discussion} \label{sec:disc} 

\subsection{\psr{}} \label{subsec:disc_psr} 

At first blush, \psr{} is an unremarkable member of the growing
population of $\gamma$-ray MSPs.  The orbital characteristics suggest
an epoch of Roche lobe overflow from an evolved 1--2$\,\msun$
companion which later became the $\sim$0.2$\,\msun$ He white dwarf
currently inferred in the system \citep[e.g.][]{tauris&savonije_1999}.
However, as we discuss below, \psr's light curve is challenging to
explain with simple geometric models that tie the radio and $\gamma$
emission to particular regions of the magnetosphere.
%Likewise, the $\gamma$-ray
%spectrum, with $\Gamma=0.9$ and a cutoff energy of $1.6$ GeV, is
%spectrum is
%typical of MSPs discovered to date \citep{msp_pop,psr_cat}.  

Remarkably, a simple picture of $\gamma$-ray emission arising in the
outer magnetosphere appears to describe the majority of both young
\citep[though see][]{romani+_2011_subl} {\em and} recycled pulsars
\citep[e.g.][]{venter+_2009}.  On the other hand, while radio emission
in young pulsars and some MSPs appears to come from lower altitudes,
several MSPs have phase-aligned $\gamma$ and radio peaks
\citep[e.g.][]{abdo+_j0034}, implying a joint emission site at high
altitude.  These observations raise the prospect of MSPs whose radio
emission combines ``traditional'' polar cap emission with a
higher-altitude component. 

To determine which scenario best describes the emission pattern of
\psr{}, we performed joint fits to the radio and $\gamma$-ray data
following the method of \citet{johnson_thesis}.  We modelled the radio
as a simple ``cone'' component \citep{story+_2007} and the $\gamma$
rays using both the ``two-pole caustic'' (TPC) model of \citet[][N.B.
we adopt a larger maximum cylindrical radius of 0.95 of the light
cylinder]{dyks&rudak_2003} and the ``outer gap'' (OG) model
\citep{chr_1986,romani&yadigaroglu_1995}.  The angle between the
pulsar spin axis and the line-of-sight, $\zeta$, is unknown {\it a
priori}, but if the radio emission arises from low altitudes, then the
presence of two peaks separated by roughly half of a rotation
indicates the inclination of the magnetic axis from the spin axis
($\alpha$) cannot be small.  The best-fit light curves for the two
$\gamma$-ray models appear in Figure~\ref{fig:lcmodel}.  The low
$\alpha=26\arcdeg$ preferred for the TPC model cannot produce the
radio interpulse and produces too much off-pulse $\gamma$-ray
emission.  The OG model, with a best fit at $\alpha=90\arcdeg$
(orthogonal rotator), produces two radio peaks but fails to produce
two clear $\gamma$ peaks and the radio interpulse morphology.

Since neither model faithfully reproduces the observed light curves,
we propose that one or both of the radio peaks may originate at
higher altitude.  Emission over a range of altitudes would also
explain the low observed linear polarization of \psr{}
\citep[e.g.][]{dyks_harding_rudak_2004} and admit smaller $\alpha$
values.  Alternatively, the failure of geometric models may indicate
that the physical details of the MSP magnetosphere---e.g.,
currents, multipolar fields, and plasma loading of the field
lines---and the radio and $\gamma$ emission mechanisms play an
important role.

\subsection{Survey Implications} \label{subsec:disc_surv}

With 6/14 pointings resulting in unbiased MSP detections, this survey
was fruitful, though this high efficiency may be luck.  Indeed,
in our timing campaign, we only detected \psr{} on 28 of 35 attempts,
and several observations required prior knowledge of the ephemeris for
detection.  Thus, because of scintillation, we estimate that \psr{} is
only detectable in 2/3 observations, depending somewhat on integration
time.  Many field MSPs are in eclipsing binaries, further diminishing
the probability of detection in a single observation.

We believe the strongest factor in our search's success was the source
selection criteria.  E.g., \citet{ransom+_3msps} selected nonvariable
LAT sources (but including no spectral shape information) and found
MSPs in 3/25 pointings.  But when considering only targets with
$|b|>5\arcdeg$, their efficiency jumps to 3/8.  \citet{keith+_2011}
employed some spectral information and detected MSPs unbiasedly in
1/11 pointings (1/4 for $|b|>5\arcdeg$).  A short list of 7
spectrally-exceptional candidates drawn up for a Nan\c{c}ay radio
telescope survey included 5 MSPs, three of which were discovered from
Nan\c{c}ay observations \citep{cognard+_2011,guillemot+_2011}.  And,
\cite{hessels+_sardegna} used nearly identical selection criteria as
this work in a Green Bank Telescope survey at 350 MHz and found MSPs
in 13/49 pointings, all but one at $|b|>5\arcdeg$.%  The use of all
the
%information the LAT can provide is clearly a strong factor in finding
%pulsars!

Targeting pulsar-like sources is clearly efficient.  However, new LAT
sources will lie at the sensitivity threshold and the limited
statistics will not admit classification of spectral shape and
variability.  Nonetheless, we argue that future radio searches should
omit detailed ranking schemes and target {\em any} LAT source off the
Galactic plane that has not been associated with a known blazar.

We believe this substantially increased target list (and telescope
time) is justified.  In the 2FGL source catalog \citep{2FGL}, there
are $\sim$350 unassociated sources more than 5\arcdeg\ off the
Galactic plane.  MSPs account for roughly 5\% of associated LAT
sources, so 15--20 MSPs may remain to be found in these unassociated
2FGL sources.  With only 350 positions to monitor, multiple deep
observations can ameliorate the confounding effects of scintillation
and eclipses while requiring a fraction of the time of a comparably
complete all-sky survey.  By comparison, the otherwise prodigious Parkes
multibeam survey found ``only'' 17 MSPs in over 40,000 pointings
\citep[e.g.][]{faulkner+_parkes_mb_msps}.  In addition to the
intrinsic interest in discovering new MSPs, by thoroughly searching
all such LAT sources, we will gain confidence that the MSPs we have
found can truly provide a volume-limited sample of the energetic,
$\gamma$-loud MSPs, an invaluable step in the study of the MSP
population of the Milky Way.
%  Given the steep radio spectra of MSPs and
%large telescope beams, low frequency searches may be especially
%fruitful. 

\acknowledgements
We thank the marvelous staff at Parkes that make it such a wonderful
research facility, Willem van Straten for help with PSRCHIVE, and
Jules Halpern for help with optical analysis.  The Parkes
Observatory is part of the Australia Telescope, which is funded by the
Commonwealth of Australia for operation as a National Facility managed
by CSIRO.

% The \fermi\ LAT Collaboration acknowledges generous ongoing support
% from a number of agencies and institutes that have supported both the
% development and the operation of the LAT as well as scientific data
% analysis.  These include the National Aeronautics and Space Administration
% and the Department of Energy in the United States, the Commissariat \`a
% l'Energie Atomique and the Centre National de la Recherche Scientifique /
% Institut National de Physique Nucl\'eaire et de Physique des Particules in
% France, the Agenzia Spaziale Italiana and the Istituto Nazionale di Fisica
% Nucleare in Italy, the Ministry of Education, Culture, Sports, Science
% and Technology (MEXT), High Energy Accelerator Research Organization
% (KEK) and Japan Aerospace Exploration Agency (JAXA) in Japan, and the
% K.~A.~Wallenberg Foundation, the Swedish Research Council and the Swedish
% National Space Board in Sweden.

The $Fermi$ LAT Collaboration acknowledges support from a number of
agencies and institutes for both development and the operation of the
LAT as well as scientific data analysis. These include NASA and DOE in
the United States, CEA/Irfu and IN2P3/CNRS in France, ASI and INFN in
Italy, MEXT, KEK, and JAXA in Japan, and the K.~A.~Wallenberg
Foundation, the Swedish Research Council and the National Space Board
in Sweden. Additional support from INAF in Italy and CNES in France
for science analysis during the operations phase is also gratefully
acknowledged.

Support for this work was provided by NASA through Einstein
Postdoctoral Fellowship Award Number PF0-110073 issued by the Chandra
X-ray Observatory Center, which is operated by the SAO for and on
behalf of NASA under contract NAS8-03060.

% Support for this work was provided by the National Aeronautics and
% Space Administration through Einstein Postdoctoral Fellowship Award
% Number PF0-110073 issued by the Chandra X-ray Observatory Center,
% which is operated by the Smithsonian Astrophysical Observatory for and
% on behalf of the National Aeronautics Space Administration under
% contract NAS8-03060.

{\em Facilities:}  \facility{{\it Fermi} (LAT)}, \facility{Parkes (PMDAQ,
PDFB3)}

\bibliographystyle{apj}

\begin{thebibliography}{35}
\expandafter\ifx\csname natexlab\endcsname\relax\def\natexlab#1{#1}\fi

\bibitem[{{Abdo} {et~al.}(2009{\natexlab{a}}){Abdo}, {Ackermann}, {Ajello},
  {Atwood}, {Axelsson}, {Baldini}, {Ballet}, {Barbiellini}, {Baring},
  {Bastieri}, {Baughman}, {Bechtol}, {Bellazzini}, {Berenji}, {Bignami},
  {Blandford}, {Bloom}, {Bonamente}, {Borgland}, {Bregeon}, {Brez}, {Brigida},
  {Bruel}, {Burnett}, {Caliandro}, {Cameron}, {Camilo}, {Caraveo}, {Carlson},
  {Casandjian}, {Cecchi}, {{\c C}elik}, {Charles}, {Chekhtman}, {Cheung},
  {Chiang}, {Ciprini}, {Claus}, {Cognard}, {Cohen-Tanugi}, {Cominsky},
  {Conrad}, {Corbet}, {Cutini}, {Dermer}, {Desvignes}, {de Angelis}, {de Luca},
  {de Palma}, {Digel}, {Dormody}, {do Couto e Silva}, {Drell}, {Dubois},
  {Dumora}, {Edmonds}, {Farnier}, {Favuzzi}, {Fegan}, {Focke}, {Frailis},
  {Freire}, {Fukazawa}, {Funk}, {Fusco}, {Gargano}, {Gasparrini}, {Gehrels},
  {Germani}, {Giebels}, {Giglietto}, {Giordano}, {Glanzman}, {Godfrey},
  {Grenier}, {Grondin}, {Grove}, {Guillemot}, {Guiriec}, {Hanabata}, {Harding},
  {Hayashida}, {Hays}, {Hobbs}, {Hughes}, {J{\'o}hannesson}, {Johnson},
  {Johnson}, {Johnson}, {Johnson}, {Johnston}, {Kamae}, {Katagiri}, {Kataoka},
  {Kawai}, {Kerr}, {Kn{\"o}dlseder}, {Kocian}, {Kramer}, {Kuss}, {Lande},
  {Latronico}, {Lemoine-Goumard}, {Longo}, {Loparco}, {Lott}, {Lovellette},
  {Lubrano}, {Madejski}, {Makeev}, {Manchester}, {Marelli}, {Mazziotta},
  {McConville}, {McEnery}, {McLaughlin}, {Meurer}, {Michelson}, {Mitthumsiri},
  {Mizuno}, {Moiseev}, {Monte}, {Monzani}, {Morselli}, {Moskalenko}, {Murgia},
  {Nolan}, {Norris}, {Nuss}, {Ohsugi}, {Omodei}, {Orlando}, {Ormes}, {Paneque},
  {Panetta}, {Parent}, {Pelassa}, {Pepe}, {Pesce-Rollins}, {Piron}, {Porter},
  {Rain{\`o}}, {Rando}, {Ransom}, {Ray}, {Razzano}, {Rea}, {Reimer}, {Reimer},
  {Reposeur}, {Ritz}, {Rochester}, {Rodriguez}, {Romani}, {Roth}, {Ryde},
  {Sadrozinski}, {Sanchez}, {Sander}, {Saz Parkinson}, {Scargle}, {Schalk},
  {Sgr{\`o}}, {Siskind}, {Smith}, {Smith}, {Spandre}, {Spinelli}, {Stappers},
  {Starck}, {Striani}, {Strickman}, {Suson}, {Tajima}, {Takahashi}, {Tanaka},
  {Thayer}, {Thayer}, {Theureau}, {Thompson}, {Thorsett}, {Tibaldo}, {Torres},
  {Tosti}, {Tramacere}, {Uchiyama}, {Usher}, {Van Etten}, {Vasileiou},
  {Venter}, {Vilchez}, {Vitale}, {Waite}, {Wallace}, {Wang}, {Watters}, {Webb},
  {Weltevrede}, {Winer}, {Wood}, {Ylinen}, \& {Ziegler}}]{msp_pop}
{Abdo}, A.~A., {et~al.} 2009{\natexlab{a}}, Science, 325, 848

\bibitem[{{Abdo} {et~al.}(2009{\natexlab{b}}){Abdo}, {Ackermann}, {Ajello},
  {Anderson}, {Atwood}, {Axelsson}, {Baldini}, {Ballet}, {Barbiellini},
  {Baring}, {Bastieri}, {Baughman}, {Bechtol}, {Bellazzini}, {Berenji},
  {Bignami}, {Blandford}, {Bloom}, {Bonamente}, {Borgland}, {Bregeon}, {Brez},
  {Brigida}, {Bruel}, {Burnett}, {Caliandro}, {Cameron}, {Caraveo},
  {Casandjian}, {Cecchi}, {{\c C}elik}, {Chekhtman}, {Cheung}, {Chiang},
  {Ciprini}, {Claus}, {Cohen-Tanugi}, {Conrad}, {Cutini}, {Dermer}, {de
  Angelis}, {de Luca}, {de Palma}, {Digel}, {Dormody}, {do Couto e Silva},
  {Drell}, {Dubois}, {Dumora}, {Farnier}, {Favuzzi}, {Fegan}, {Fukazawa},
  {Funk}, {Fusco}, {Gargano}, {Gasparrini}, {Gehrels}, {Germani}, {Giebels},
  {Giglietto}, {Giommi}, {Giordano}, {Glanzman}, {Godfrey}, {Grenier},
  {Grondin}, {Grove}, {Guillemot}, {Guiriec}, {Gwon}, {Hanabata}, {Harding},
  {Hayashida}, {Hays}, {Hughes}, {J{\'o}hannesson}, {Johnson}, {Johnson},
  {Johnson}, {Kamae}, {Katagiri}, {Kataoka}, {Kawai}, {Kerr}, {Kn{\"o}dlseder},
  {Kocian}, {Kuss}, {Lande}, {Latronico}, {Lemoine-Goumard}, {Longo},
  {Loparco}, {Lott}, {Lovellette}, {Lubrano}, {Madejski}, {Makeev}, {Marelli},
  {Mazziotta}, {McConville}, {McEnery}, {Meurer}, {Michelson}, {Mitthumsiri},
  {Mizuno}, {Monte}, {Monzani}, {Morselli}, {Moskalenko}, {Murgia}, {Nolan},
  {Norris}, {Nuss}, {Ohsugi}, {Omodei}, {Orlando}, {Ormes}, {Paneque},
  {Parent}, {Pelassa}, {Pepe}, {Pesce-Rollins}, {Pierbattista}, {Piron},
  {Porter}, {Primack}, {Rain{\`o}}, {Rando}, {Ray}, {Razzano}, {Rea}, {Reimer},
  {Reimer}, {Reposeur}, {Ritz}, {Rochester}, {Rodriguez}, {Romani}, {Ryde},
  {Sadrozinski}, {Sanchez}, {Sander}, {Parkinson}, {Scargle}, {Sgr{\`o}},
  {Siskind}, {Smith}, {Smith}, {Spandre}, {Spinelli}, {Starck}, {Strickman},
  {Suson}, {Tajima}, {Takahashi}, {Takahashi}, {Tanaka}, {Thayer}, {Thompson},
  {Tibaldo}, {Tibolla}, {Torres}, {Tosti}, {Tramacere}, {Uchiyama}, {Usher},
  {Van Etten}, {Vasileiou}, {Vilchez}, {Vitale}, {Waite}, {Wang}, {Watters},
  {Winer}, {Wolff}, {Wood}, {Ylinen}, \& {Ziegler}}]{blind_search_16}
---. 2009{\natexlab{b}}, Science, 325, 840

\bibitem[{{Abdo} {et~al.}(2010{\natexlab{a}}){Abdo}, {Ackermann}, {Ajello},
  {Allafort}, {Baldini}, {Ballet}, {Barbiellini}, {Bastieri}, {Bechtol},
  {Bellazzini}, {Berenji}, {Blandford}, {Bloom}, {Bonamente}, {Borgland},
  {Bouvier}, {Bregeon}, {Brez}, {Brigida}, {Bruel}, {Burnett}, {Buson},
  {Caliandro}, {Cameron}, {Camilo}, {Caraveo}, {Carrigan}, {Casandjian},
  {Cecchi}, {{\c C}elik}, {Chekhtman}, {Cheung}, {Chiang}, {Ciprini}, {Claus},
  {Cognard}, {Cohen-Tanugi}, {Conrad}, {Corbet}, {DeCesar}, {Dermer},
  {Desvignes}, {de Angelis}, {de Palma}, {Digel}, {Dormody}, {Silva}, {Drell},
  {Dubois}, {Dumora}, {Espinoza}, {Farnier}, {Favuzzi}, {Fegan}, {Focke},
  {Frailis}, {Freire}, {Fukazawa}, {Funk}, {Fusco}, {Gargano}, {Gasparrini},
  {Gehrels}, {Germani}, {Giavitto}, {Giglietto}, {Giordano}, {Glanzman},
  {Godfrey}, {Grenier}, {Grondin}, {Grove}, {Guillemot}, {Guiriec}, {Hadasch},
  {Harding}, {Hays}, {Hobbs}, {Horan}, {Hughes}, {J{\'o}hannesson}, {Johnson},
  {Johnson}, {Johnson}, {Johnston}, {Kamae}, {Katagiri}, {Kataoka}, {Kawai},
  {Kerr}, {Kn{\"o}dlseder}, {Kramer}, {Kuss}, {Lande}, {Latronico},
  {Lemoine-Goumard}, {Llena Garde}, {Longo}, {Loparco}, {Lott}, {Lovellette},
  {Lubrano}, {Lyne}, {Makeev}, {Manchester}, {Marelli}, {Mazziotta},
  {McConville}, {McEnery}, {McGlynn}, {Meurer}, {Michelson}, {Mitthumsiri},
  {Mizuno}, {Moiseev}, {Monte}, {Monzani}, {Morselli}, {Moskalenko}, {Murgia},
  {Nolan}, {Norris}, {Noutsos}, {Nuss}, {Ohsugi}, {Omodei}, {Orlando}, {Ormes},
  {Ozaki}, {Paneque}, {Panetta}, {Parent}, {Pelassa}, {Pepe}, {Pesce-Rollins},
  {Pierbattista}, {Piron}, {Porter}, {Rain{\`o}}, {Rando}, {Ransom}, {Razzano},
  {Reimer}, {Reimer}, {Reposeur}, {Ripken}, {Ritz}, {Rochester}, {Rodriguez},
  {Romani}, {Roth}, {Ryde}, {Sadrozinski}, {Sander}, {Saz Parkinson},
  {Scargle}, {Sgr{\`o}}, {Siskind}, {Smith}, {Smith}, {Spandre}, {Spinelli},
  {Stappers}, {Starck}, {Strickman}, {Suson}, {Takahashi}, {Tanaka}, {Thayer},
  {Thayer}, {Theureau}, {Thompson}, {Thorsett}, {Tibaldo}, {Torres}, {Tosti},
  {Tramacere}, {Usher}, {Van Etten}, {Vasileiou}, {Venter}, {Vilchez},
  {Vitale}, {Waite}, {Wallace}, {Wang}, {Weltevrede}, {Winer}, {Wood},
  {Ylinen}, \& {Ziegler}}]{abdo+_j0034}
---. 2010{\natexlab{a}}, \apj, 712, 957

\bibitem[{{Abdo} {et~al.}(2010{\natexlab{b}}){Abdo}, {Ackermann}, {Ajello},
  {Allafort}, {Antolini}, {Atwood}, {Axelsson}, {Baldini}, {Ballet},
  {Barbiellini}, {Bastieri}, {Baughman}, {Bechtol}, {Bellazzini}, {Belli},
  {Berenji}, {Bisello}, {Blandford}, {Bloom}, {Bonamente}, {Bonnell},
  {Borgland}, {Bouvier}, {Bregeon}, {Brez}, {Brigida}, {Bruel}, {Burnett},
  {Busetto}, {Buson}, {Caliandro}, {Cameron}, {Campana}, {Canadas}, {Caraveo},
  {Carrigan}, {Casandjian}, {Cavazzuti}, {Ceccanti}, {Cecchi}, {{\c C}elik},
  {Charles}, {Chekhtman}, {Cheung}, {Chiang}, {Cillis}, {Ciprini}, {Claus},
  {Cohen-Tanugi}, {Conrad}, {Corbet}, {Davis}, {DeKlotz}, {den Hartog},
  {Dermer}, {de Angelis}, {de Luca}, {de Palma}, {Digel}, {Dormody}, {Silva},
  {Drell}, {Dubois}, {Dumora}, {Fabiani}, {Farnier}, {Favuzzi}, {Fegan},
  {Ferrara}, {Focke}, {Fortin}, {Frailis}, {Fukazawa}, {Funk}, {Fusco},
  {Gargano}, {Gasparrini}, {Gehrels}, {Germani}, {Giavitto}, {Giebels},
  {Giglietto}, {Giommi}, {Giordano}, {Giroletti}, {Glanzman}, {Godfrey},
  {Grenier}, {Grondin}, {Grove}, {Guillemot}, {Guiriec}, {Gustafsson},
  {Hadasch}, {Hanabata}, {Harding}, {Hayashida}, {Hays}, {Healey}, {Hill},
  {Horan}, {Hughes}, {Iafrate}, {J{\'o}hannesson}, {Johnson}, {Johnson},
  {Johnson}, {Johnson}, {Kamae}, {Katagiri}, {Kataoka}, {Kawai}, {Kerr},
  {Kn{\"o}dlseder}, {Kocevski}, {Kuss}, {Lande}, {Landriu}, {Latronico}, {Lee},
  {Lemoine-Goumard}, {Lionetto}, {Llena Garde}, {Longo}, {Loparco}, {Lott},
  {Lovellette}, {Lubrano}, {Madejski}, {Makeev}, {Marangelli}, {Marelli},
  {Massaro}, {Mazziotta}, {McConville}, {McEnery}, {Michelson}, {Minuti},
  {Mitthumsiri}, {Mizuno}, {Moiseev}, {Mongelli}, {Monte}, {Monzani},
  {Moretti}, {Morselli}, {Moskalenko}, {Murgia}, {Nakajima}, {Nakamori},
  {Naumann-Godo}, {Nolan}, {Norris}, {Nuss}, {Ohno}, {Ohsugi}, {Omodei},
  {Orlando}, {Ormes}, {Ozaki}, {Paccagnella}, {Paneque}, {Panetta}, {Parent},
  {Pelassa}, {Pepe}, {Pesce-Rollins}, {Pinchera}, {Piron}, {Porter}, {Poupard},
  {Rain{\`o}}, {Rando}, {Ray}, {Razzano}, {Razzaque}, {Rea}, {Reimer},
  {Reimer}, {Reposeur}, {Ripken}, {Ritz}, {Rochester}, {Rodriguez}, {Romani},
  {Roth}, {Sadrozinski}, {Salvetti}, {Sanchez}, {Sander}, {Saz Parkinson},
  {Scargle}, {Schalk}, {Scolieri}, {Sgr{\`o}}, {Shaw}, {Siskind}, {Smith},
  {Smith}, {Spandre}, {Spinelli}, {Starck}, {Stephens}, {Striani}, {Strickman},
  {Strong}, {Suson}, {Tajima}, {Takahashi}, {Takahashi}, {Tanaka}, {Thayer},
  {Thayer}, {Thompson}, {Tibaldo}, {Tibolla}, {Tinebra}, {Torres}, {Tosti},
  {Tramacere}, {Uchiyama}, {Usher}, {Van Etten}, {Vasileiou}, {Vilchez},
  {Vitale}, {Waite}, {Wallace}, {Wang}, {Watters}, {Winer}, {Wood}, {Yang},
  {Ylinen}, \& {Ziegler}}]{1fgl}
---. 2010{\natexlab{b}}, \apjs, 188, 405

\bibitem[{{Abdo} {et~al.}(2010{\natexlab{c}}){Abdo}, {Ackermann}, {Ajello},
  {Atwood}, {Axelsson}, {Baldini}, {Ballet}, {Barbiellini}, {Baring},
  {Bastieri}, {Baughman}, {Bechtol}, {Bellazzini}, {Berenji}, {Blandford},
  {Bloom}, {Bonamente}, {Borgland}, {Bregeon}, {Brez}, {Brigida}, {Bruel},
  {Burnett}, {Buson}, {Caliandro}, {Cameron}, {Camilo}, {Caraveo},
  {Casandjian}, {Cecchi}, {{\c C}elik}, {Charles}, {Chekhtman}, {Cheung},
  {Chiang}, {Ciprini}, {Claus}, {Cognard}, {Cohen-Tanugi}, {Cominsky},
  {Conrad}, {Corbet}, {Cutini}, {den Hartog}, {Dermer}, {de Angelis}, {de
  Luca}, {de Palma}, {Digel}, {Dormody}, {Silva}, {Drell}, {Dubois}, {Dumora},
  {Espinoza}, {Farnier}, {Favuzzi}, {Fegan}, {Ferrara}, {Focke}, {Fortin},
  {Frailis}, {Freire}, {Fukazawa}, {Funk}, {Fusco}, {Gargano}, {Gasparrini},
  {Gehrels}, {Germani}, {Giavitto}, {Giebels}, {Giglietto}, {Giommi},
  {Giordano}, {Glanzman}, {Godfrey}, {Gotthelf}, {Grenier}, {Grondin}, {Grove},
  {Guillemot}, {Guiriec}, {Gwon}, {Hanabata}, {Harding}, {Hayashida}, {Hays},
  {Hughes}, {Jackson}, {J{\'o}hannesson}, {Johnson}, {Johnson}, {Johnson},
  {Johnson}, {Johnston}, {Kamae}, {Kanbach}, {Kaspi}, {Katagiri}, {Kataoka},
  {Kawai}, {Kerr}, {Kn{\"o}dlseder}, {Kocian}, {Kramer}, {Kuss}, {Lande},
  {Latronico}, {Lemoine-Goumard}, {Livingstone}, {Longo}, {Loparco}, {Lott},
  {Lovellette}, {Lubrano}, {Lyne}, {Madejski}, {Makeev}, {Manchester},
  {Marelli}, {Mazziotta}, {McConville}, {McEnery}, {McGlynn}, {Meurer},
  {Michelson}, {Mineo}, {Mitthumsiri}, {Mizuno}, {Moiseev}, {Monte}, {Monzani},
  {Morselli}, {Moskalenko}, {Murgia}, {Nakamori}, {Nolan}, {Norris}, {Noutsos},
  {Nuss}, {Ohsugi}, {Omodei}, {Orlando}, {Ormes}, {Ozaki}, {Paneque},
  {Panetta}, {Parent}, {Pelassa}, {Pepe}, {Pesce-Rollins}, {Piron}, {Porter},
  {Rain{\`o}}, {Rando}, {Ransom}, {Ray}, {Razzano}, {Rea}, {Reimer}, {Reimer},
  {Reposeur}, {Ritz}, {Rodriguez}, {Romani}, {Roth}, {Ryde}, {Sadrozinski},
  {Sanchez}, {Sander}, {Saz Parkinson}, {Scargle}, {Schalk}, {Sellerholm},
  {Sgr{\`o}}, {Siskind}, {Smith}, {Smith}, {Spandre}, {Spinelli}, {Stappers},
  {Starck}, {Striani}, {Strickman}, {Strong}, {Suson}, {Tajima}, {Takahashi},
  {Takahashi}, {Tanaka}, {Thayer}, {Thayer}, {Theureau}, {Thompson},
  {Thorsett}, {Tibaldo}, {Tibolla}, {Torres}, {Tosti}, {Tramacere}, {Uchiyama},
  {Usher}, {Van Etten}, {Vasileiou}, {Venter}, {Vilchez}, {Vitale}, {Waite},
  {Wang}, {Wang}, {Watters}, {Weltevrede}, {Winer}, {Wood}, {Ylinen}, \&
  {Ziegler}}]{psr_cat}
---. 2010{\natexlab{c}}, \apjs, 187, 460

\bibitem[{{Abdo} {et~al.}(2011){Abdo}, {Ackermann}, {Ajello}, {Allafort},
  {Antolini}, {Atwood}, {Axelsson}, {Baldini}, {Ballet}, {Barbiellini},
  {Bastieri}, {Baughman}, {Bechtol}, {Bellazzini}, {Belli}, {Berenji},
  {Bisello}, {Blandford}, {Bloom}, {Bonamente}, {Bonnell}, {Borgland},
  {Bouvier}, {Bregeon}, {Brez}, {Brigida}, {Bruel}, {Burnett}, {Busetto},
  {Buson}, {Caliandro}, {Cameron}, {Campana}, {Canadas}, {Caraveo}, {Carrigan},
  {Casandjian}, {Cavazzuti}, {Ceccanti}, {Cecchi}, {{\c C}elik}, {Charles},
  {Chekhtman}, {Cheung}, {Chiang}, {Cillis}, {Ciprini}, {Claus},
  {Cohen-Tanugi}, {Conrad}, {Corbet}, {Davis}, {DeKlotz}, {den Hartog},
  {Dermer}, {de Angelis}, {de Luca}, {de Palma}, {Digel}, {Dormody}, {Silva},
  {Drell}, {Dubois}, {Dumora}, {Fabiani}, {Farnier}, {Favuzzi}, {Fegan},
  {Ferrara}, {Focke}, {Fortin}, {Frailis}, {Fukazawa}, {Funk}, {Fusco},
  {Gargano}, {Gasparrini}, {Gehrels}, {Germani}, {Giavitto}, {Giebels},
  {Giglietto}, {Giommi}, {Giordano}, {Giroletti}, {Glanzman}, {Godfrey},
  {Grenier}, {Grondin}, {Grove}, {Guillemot}, {Guiriec}, {Gustafsson},
  {Hadasch}, {Hanabata}, {Harding}, {Hayashida}, {Hays}, {Healey}, {Hill},
  {Horan}, {Hughes}, {Iafrate}, {J{\'o}hannesson}, {Johnson}, {Johnson},
  {Johnson}, {Johnson}, {Kamae}, {Katagiri}, {Kataoka}, {Kawai}, {Kerr},
  {Kn{\"o}dlseder}, {Kocevski}, {Kuss}, {Lande}, {Landriu}, {Latronico}, {Lee},
  {Lemoine-Goumard}, {Lionetto}, {Llena Garde}, {Longo}, {Loparco}, {Lott},
  {Lovellette}, {Lubrano}, {Madejski}, {Makeev}, {Marangelli}, {Marelli},
  {Massaro}, {Mazziotta}, {McConville}, {McEnery}, {Michelson}, {Minuti},
  {Mitthumsiri}, {Mizuno}, {Moiseev}, {Mongelli}, {Monte}, {Monzani},
  {Moretti}, {Morselli}, {Moskalenko}, {Murgia}, {Nakajima}, {Nakamori},
  {Naumann-Godo}, {Nolan}, {Norris}, {Nuss}, {Ohno}, {Ohsugi}, {Omodei},
  {Orlando}, {Ormes}, {Ozaki}, {Paccagnella}, {Paneque}, {Panetta}, {Parent},
  {Pelassa}, {Pepe}, {Pesce-Rollins}, {Pinchera}, {Piron}, {Porter}, {Poupard},
  {Rain{\`o}}, {Rando}, {Ray}, {Razzano}, {Razzaque}, {Rea}, {Reimer},
  {Reimer}, {Reposeur}, {Ripken}, {Ritz}, {Rochester}, {Rodriguez}, {Romani},
  {Roth}, {Sadrozinski}, {Salvetti}, {Sanchez}, {Sander}, {Saz Parkinson},
  {Scargle}, {Schalk}, {Scolieri}, {Sgr{\`o}}, {Shaw}, {Siskind}, {Smith},
  {Smith}, {Spandre}, {Spinelli}, {Starck}, {Stephens}, {Striani}, {Strickman},
  {Strong}, {Suson}, {Tajima}, {Takahashi}, {Takahashi}, {Tanaka}, {Thayer},
  {Thayer}, {Thompson}, {Tibaldo}, {Tibolla}, {Tinebra}, {Torres}, {Tosti},
  {Tramacere}, {Uchiyama}, {Usher}, {Van Etten}, {Vasileiou}, {Vilchez},
  {Vitale}, {Waite}, {Wallace}, {Wang}, {Watters}, {Winer}, {Wood}, {Yang},
  {Ylinen}, \& {Ziegler}}]{2FGL}
---. 2011, arXiv:1108.1435

\bibitem[{{Atwood} {et~al.}(2009){Atwood}, {Abdo}, {Ackermann}, {Althouse},
  {Anderson}, {Axelsson}, {Baldini}, {Ballet}, {Band}, {Barbiellini},
  {Bartelt}, {Bastieri}, {Baughman}, {Bechtol}, {B{\'e}d{\'e}r{\`e}de},
  {Bellardi}, {Bellazzini}, {Berenji}, {Bignami}, {Bisello}, {Bissaldi},
  {Blandford}, {Bloom}, {Bogart}, {Bonamente}, {Bonnell}, {Borgland},
  {Bouvier}, {Bregeon}, {Brez}, {Brigida}, {Bruel}, {Burnett}, {Busetto},
  {Caliandro}, {Cameron}, {Caraveo}, {Carius}, {Carlson}, {Casandjian},
  {Cavazzuti}, {Ceccanti}, {Cecchi}, {Charles}, {Chekhtman}, {Cheung},
  {Chiang}, {Chipaux}, {Cillis}, {Ciprini}, {Claus}, {Cohen-Tanugi},
  {Condamoor}, {Conrad}, {Corbet}, {Corucci}, {Costamante}, {Cutini}, {Davis},
  {Decotigny}, {DeKlotz}, {Dermer}, {de Angelis}, {Digel}, {do Couto e Silva},
  {Drell}, {Dubois}, {Dumora}, {Edmonds}, {Fabiani}, {Farnier}, {Favuzzi},
  {Flath}, {Fleury}, {Focke}, {Funk}, {Fusco}, {Gargano}, {Gasparrini},
  {Gehrels}, {Gentit}, {Germani}, {Giebels}, {Giglietto}, {Giommi}, {Giordano},
  {Glanzman}, {Godfrey}, {Grenier}, {Grondin}, {Grove}, {Guillemot}, {Guiriec},
  {Haller}, {Harding}, {Hart}, {Hays}, {Healey}, {Hirayama}, {Hjalmarsdotter},
  {Horn}, {Hughes}, {J{\'o}hannesson}, {Johansson}, {Johnson}, {Johnson},
  {Johnson}, {Johnson}, {Kamae}, {Katagiri}, {Kataoka}, {Kavelaars}, {Kawai},
  {Kelly}, {Kerr}, {Klamra}, {Kn{\"o}dlseder}, {Kocian}, {Komin}, {Kuehn},
  {Kuss}, {Landriu}, {Latronico}, {Lee}, {Lee}, {Lemoine-Goumard}, {Lionetto},
  {Longo}, {Loparco}, {Lott}, {Lovellette}, {Lubrano}, {Madejski}, {Makeev},
  {Marangelli}, {Massai}, {Mazziotta}, {McEnery}, {Menon}, {Meurer},
  {Michelson}, {Minuti}, {Mirizzi}, {Mitthumsiri}, {Mizuno}, {Moiseev},
  {Monte}, {Monzani}, {Moretti}, {Morselli}, {Moskalenko}, {Murgia},
  {Nakamori}, {Nishino}, {Nolan}, {Norris}, {Nuss}, {Ohno}, {Ohsugi}, {Omodei},
  {Orlando}, {Ormes}, {Paccagnella}, {Paneque}, {Panetta}, {Parent}, {Pearce},
  {Pepe}, {Perazzo}, {Pesce-Rollins}, {Picozza}, {Pieri}, {Pinchera}, {Piron},
  {Porter}, {Poupard}, {Rain{\`o}}, {Rando}, {Rapposelli}, {Razzano}, {Reimer},
  {Reimer}, {Reposeur}, {Reyes}, {Ritz}, {Rochester}, {Rodriguez}, {Romani},
  {Roth}, {Russell}, {Ryde}, {Sabatini}, {Sadrozinski}, {Sanchez}, {Sander},
  {Sapozhnikov}, {Parkinson}, {Scargle}, {Schalk}, {Scolieri}, {Sgr{\`o}},
  {Share}, {Shaw}, {Shimokawabe}, {Shrader}, {Sierpowska-Bartosik}, {Siskind},
  {Smith}, {Smith}, {Spandre}, {Spinelli}, {Starck}, {Stephens}, {Strickman},
  {Strong}, {Suson}, {Tajima}, {Takahashi}, {Takahashi}, {Tanaka}, {Tenze},
  {Tether}, {Thayer}, {Thayer}, {Thompson}, {Tibaldo}, {Tibolla}, {Torres},
  {Tosti}, {Tramacere}, {Turri}, {Usher}, {Vilchez}, {Vitale}, {Wang},
  {Watters}, {Winer}, {Wood}, {Ylinen}, \& {Ziegler}}]{lat_instrument}
{Atwood}, W.~B., {et~al.} 2009, \apj, 697, 1071

\bibitem[{{Camilo} {et~al.}(1994){Camilo}, {Thorsett}, \&
  {Kulkarni}}]{camilo+_1994}
{Camilo}, F., {Thorsett}, S.~E., \& {Kulkarni}, S.~R. 1994, \apjl, 421, L15

\bibitem[{{Cheng} {et~al.}(1986){Cheng}, {Ho}, \& {Ruderman}}]{chr_1986}
{Cheng}, K.~S., {Ho}, C., \& {Ruderman}, M. 1986, \apj, 300, 500

\bibitem[{{Cognard} {et~al.}(2011){Cognard}, {Guillemot}, {Johnson}, {Smith},
  {Venter}, {Harding}, {Wolff}, {Cheung}, {Donato}, {Abdo}, {Ballet}, {Camilo},
  {Desvignes}, {Dumora}, {Ferrara}, {Freire}, {Grove}, {Johnston}, {Keith},
  {Kramer}, {Lyne}, {Michelson}, {Parent}, {Ransom}, {Ray}, {Romani}, {Saz
  Parkinson}, {Stappers}, {Theureau}, {Thompson}, {Weltevrede}, \&
  {Wood}}]{cognard+_2011}
{Cognard}, I., {et~al.} 2011, \apj, 732, 47

\bibitem[{{Cordes} \& {Lazio}(2002)}]{cl02}
{Cordes}, J.~M., \& {Lazio}, T.~J.~W. 2002, ArXiv Astrophysics e-prints
  arXiv:astro-ph/0207156

\bibitem[{{de Jager} {et~al.}(1989){de Jager}, {Raubenheimer}, \&
  {Swanepoel}}]{dejager_1}
{de Jager}, O.~C., {Raubenheimer}, B.~C., \& {Swanepoel}, J.~W.~H. 1989, \aap,
  221, 180

\bibitem[{{Dyks} {et~al.}(2004){Dyks}, {Harding}, \&
  {Rudak}}]{dyks_harding_rudak_2004}
{Dyks}, J., {Harding}, A.~K., \& {Rudak}, B. 2004, \apj, 606, 1125

\bibitem[{{Dyks} \& {Rudak}(2003)}]{dyks&rudak_2003}
{Dyks}, J., \& {Rudak}, B. 2003, \apj, 598, 1201

\bibitem[{{Faulkner} {et~al.}(2004){Faulkner}, {Stairs}, {Kramer}, {Lyne},
  {Hobbs}, {Possenti}, {Lorimer}, {Manchester}, {McLaughlin}, {D'Amico},
  {Camilo}, \& {Burgay}}]{faulkner+_parkes_mb_msps}
{Faulkner}, A.~J., {et~al.} 2004, \mnras, 355, 147

\bibitem[{{Guillemot} {et~al.}(2011){Guillemot}, {Freire}, {Cognard},
  {Johnson}, {Takahashi}, {Kataoka}, {Desvignes}, {Camilo}, {Ferrara},
  {Harding}, {Janssen}, {Keith}, {Kerr}, {Kramer}, {Parent}, {Ransom}, {Ray},
  {Saz Parkinson}, {Smith}, {Stappers}, \& {Theureau}}]{guillemot+_2011}
{Guillemot}, L., {et~al.} 2011, \mnras, submitted

\bibitem[{{Hessels} {et~al.}(2011){Hessels}, {Roberts}, {McLaughlin}, {Ray},
  {Bangale}, {Ransom}, {Kerr}, {Camilo}, {DeCesar}, \& {Pulsar Search
  Consortium}}]{hessels+_sardegna}
{Hessels}, J.~W.~T., {et~al.} 2011, arXiv:1101.1742

\bibitem[{{Hotan} {et~al.}(2004){Hotan}, {van Straten}, \&
  {Manchester}}]{hotan&manchester_2004}
{Hotan}, A.~W., {van Straten}, W., \& {Manchester}, R.~N. 2004, Publications of
  the Astron. Soc. of Australia, 21, 302

\bibitem[{{Johnson}(2011)}]{johnson_thesis}
{Johnson}, T. 2011, PhD Thesis, University of Maryland, arXiv:TBD

\bibitem[{{Keith} {et~al.}(2011){Keith}, {Johnston}, {Ray}, {Ferrara}, {Saz
  Parkinson}, {{\c C}elik}, {Belfiore}, {Donato}, {Cheung}, {Abdo}, {Camilo},
  {Freire}, {Guillemot}, {Harding}, {Kramer}, {Michelson}, {Ransom}, {Romani},
  {Smith}, {Thompson}, {Weltevrede}, \& {Wood}}]{keith+_2011}
{Keith}, M.~J., {et~al.} 2011, \mnras, 414, 1292

\bibitem[{{Kerr}(2011)}]{kerr_thesis}
{Kerr}, M. 2011, PhD Thesis, University of Washington, arXiv:1101.6072

\bibitem[{{Manchester} {et~al.}(2001){Manchester}, {Lyne}, {Camilo}, {Bell},
  {Kaspi}, {D'Amico}, {McKay}, {Crawford}, {Stairs}, {Possenti}, {Kramer}, \&
  {Sheppard}}]{manchester+_2001_parkes_multibeam}
{Manchester}, R.~N., {et~al.} 2001, \mnras, 328, 17

\bibitem[{{Marelli} {et~al.}(2011){Marelli}, {De Luca}, \&
  {Caraveo}}]{marelli+_2011}
{Marelli}, M., {De Luca}, A., \& {Caraveo}, P.~A. 2011, \apj, 733, 82

\bibitem[{{Nice} \& {Taylor}(1995)}]{nice&taylor_1995}
{Nice}, D.~J., \& {Taylor}, J.~H. 1995, \apj, 441, 429

\bibitem[{{Pletsch} {et~al.}(2012){Pletsch}, {Guillemot}, {Allen}, {Kramer},
  {Aulbert}, {Fehrmann}, {Ray}, {Barr}, {Belfiore}, {Camilo}, {Caraveo}, {{\c
  C}elik}, {Champion}, {Dormody}, {Eatough}, {Ferrara}, {Freire}, {Hessels},
  {Keith}, {Kerr}, {de Luca}, {Lyne}, {Marelli}, {McLaughlin}, {Parent},
  {Ransom}, {Razzano}, {Reich}, {Saz Parkinson}, {Stappers}, \&
  {Wolff}}]{pletsch+_2012}
{Pletsch}, H.~J., {et~al.} 2012, \apj, 744, 105

\bibitem[{{Ransom}(2001)}]{ransom_thesis}
{Ransom}, S.~M. 2001, PhD thesis, Harvard University

\bibitem[{{Ransom} {et~al.}(2011){Ransom}, {Ray}, {Camilo}, {Roberts}, {{\c
  C}elik}, {Wolff}, {Cheung}, {Kerr}, {Pennucci}, {DeCesar}, {Cognard}, {Lyne},
  {Stappers}, {Freire}, {Grove}, {Abdo}, {Desvignes}, {Donato}, {Ferrara},
  {Gehrels}, {Guillemot}, {Gwon}, {Harding}, {Johnston}, {Keith}, {Kramer},
  {Michelson}, {Parent}, {Saz Parkinson}, {Romani}, {Smith}, {Theureau},
  {Thompson}, {Weltevrede}, {Wood}, \& {Ziegler}}]{ransom+_3msps}
{Ransom}, S.~M., {et~al.} 2011, \apjl, 727, L16

\bibitem[{{Romani} {et~al.}(2011){Romani}, {Kerr}, {Craig}, {Johnston},
  {Cognard}, \& {Smith}}]{romani+_2011_subl}
{Romani}, R.~W., {Kerr}, M., {Craig}, H.~A., {Johnston}, S., {Cognard}, I., \&
  {Smith}, D.~A. 2011, \apj, 738, 114

\bibitem[{{Romani} \& {Yadigaroglu}(1995)}]{romani&yadigaroglu_1995}
{Romani}, R.~W., \& {Yadigaroglu}, I.-A. 1995, \apj, 438, 314

\bibitem[{{Shklovskii}(1970)}]{shklovskii}
{Shklovskii}, I.~S. 1970, \sovast, 13, 562

\bibitem[{{Smith} {et~al.}(2008){Smith}, {Guillemot}, {Camilo}, {Cognard},
  {Dumora}, {Espinoza}, {Freire}, {Gotthelf}, {Harding}, {Hobbs}, {Johnston},
  {Kaspi}, {Kramer}, {Livingstone}, {Lyne}, {Manchester}, {Marshall},
  {McLaughlin}, {Noutsos}, {Ransom}, {Roberts}, {Romani}, {Stappers},
  {Theureau}, {Thompson}, {Thorsett}, {Wang}, \& {Weltevrede}}]{smith_timing}
{Smith}, D.~A., {et~al.} 2008, \aap, 492, 923

\bibitem[{{Story} {et~al.}(2007){Story}, {Gonthier}, \&
  {Harding}}]{story+_2007}
{Story}, S.~A., {Gonthier}, P.~L., \& {Harding}, A.~K. 2007, \apj, 671, 713

\bibitem[{{Tauris} \& {Savonije}(1999)}]{tauris&savonije_1999}
{Tauris}, T.~M., \& {Savonije}, G.~J. 1999, \aap, 350, 928

\bibitem[{{Venter} {et~al.}(2009){Venter}, {Harding}, \&
  {Guillemot}}]{venter+_2009}
{Venter}, C., {Harding}, A.~K., \& {Guillemot}, L. 2009, \apj, 707, 800

\bibitem[{{Watters} {et~al.}(2009){Watters}, {Romani}, {Weltevrede}, \&
  {Johnston}}]{watters+_2009}
{Watters}, K.~P., {Romani}, R.~W., {Weltevrede}, P., \& {Johnston}, S. 2009,
  \apj, 695, 1289

\end{thebibliography}

\end{document}